\documentclass[acmsmall,nonacm,11pt]{acmart}
\makeatletter\let\@authorsaddresses\@empty\makeatother
\usepackage[noframes,nonumbers]{ffcode}
\makeatletter
\lst@AddToHook{DisplayStyle}{\let\@vspace\@vspace@orig\let\@vspacer\@vspacer@orig}
\makeatother
\usepackage{natbib}
\usepackage{doi}
\usepackage{eolang}
\usepackage{href-ul}
\usepackage[capitalize]{cleveref}
\usepackage{to-be-determined}
\usepackage[utf8]{inputenc}
\usepackage[T2A,T1]{fontenc}
\usepackage[russian,english]{babel}
  \DeclareFontFamilySubstitution{T2A}{LinuxBiolinumT-TLF}{LibertinusSans-TLF}
  
\DeclareFontFamily{U}{stmry}{}
\DeclareFontShape{U}{stmry}{m}{n}{%
  <5><6><7><8><9><10>gen*stmary%
  <5.5>stmary5%
  <10.95><12><14.4><17.28><20.74><24.88>stmary10%
}{}
\usepackage{csquotes}
\usepackage{url}

\title{On the Origin of Objects}
\subtitle{By Means of Careful Selection}

\author{Yegor Bugayenko}
\orcid{0000-0001-6370-0678}
\email{yegor256@gmail.com}
\affiliation{
  \institution{Huawei}
  \country{Russia}
  \city{Moscow}
  }

\author{Maxim Trunnikov}
\orcid{0000-0001-8545-4797}
\email{mtrunnikov@gmail.com}
\affiliation{
  \institution{Huawei}
  \country{Russia}
  \city{Moscow}
  }


\newcommand\aff[1]{\ff{\textcolor{gray}{\(\star\)}#1}}
\newcommand\deff[1]{\ff{\textcolor{blue!50!black}{\textbf{#1}}}}
\newcommand\adeff[1]{\aff{\textcolor{blue!50!black}{\textbf{#1}}}}
\newcommand\eohex[1]{\ff{#1}}

\begin{document}

\begin{abstract}
We introduce a taxonomy of objects for the \eolang{} programming language.
This taxonomy is designed with a few principles in mind:
  non-redundancy and simplicity.
The taxonomy is supposed to be used
  as a navigation map by \eolang{} programmers.
It may also be helpful as a guideline for designers
  of other object-oriented languages or libraries for them.
\end{abstract}

\maketitle

\section{Introduction}

Each object-oriented programming language offers its own set of objects,
  classes, types, functions, constants, traits, templates, and so on,
  which programmers can use off-the-shelf to model their specific use cases:
  ``Development Kit'' in Java~\citep{jdk2024,jdk8},
  ``Standard Library'' in
    C++~\citep{cpp2024,josuttis2012cpp},
    Python~\citep{python2024,hellmann2017python},
    Swift~\citep{swift2024,deitel2015swift},
    Rust~\citep{rust2024,blandy2021rust},
    and
    Ruby,
  ``.NET Standard'' in C\#~\citep{net2023,abrams2005net},
  and ``Built-in Objects'' in JavaScript~\citep{js2024,crockford2008js}.

Each standard library (SL) freely defines its own principles of abstraction.
For example, to get the absolute value of a numeric object \ff{x} in Java,
  one has to call a static method of another class \ff{Math.abs(x)},
  while in Ruby it would be a property of the same object \ff{x.abs}.
For example, in Java, an attempt to get a value from a hash map by
  a key will produce \ff{null} in case of its absence,
  while in C++ it will return an empty iterator.
There are many similar examples demonstrating
  the absence of a common ground between SLs.

Earlier, \citet{booch1990design} suggested their own components for OOP.
We suggest our own taxonomy of objects for \eolang{}\footnote{%
  \url{https://www.eolang.org}, 0.61.3},
  a strictly formal~\citep{kudasov2022formalizing} object-oriented
  programming language with an intentionally reduced
  feature set~\citep{bugayenko2021eolang}.
All objects are grouped as such:

\begin{itemize}
    \item Bytes (\cref{sec:bytes})
    \item Boolean (\cref{sec:boolean})
    \item Number (\cref{sec:number})
    \item String (\cref{sec:string})
    \item Tuple (\cref{sec:tuple})
    \item Error handling (\cref{sec:errors})
    \item Flow control abstractions (\cref{sec:flow})
    \item Non-standard bit-width numbers (\cref{sec:digits})
    \item Memory abstractions (\cref{sec:memory})
    \item Math functions and algorithms (\cref{sec:math})
    \item Text manipulations (\cref{sec:strings})
    \item Structs (\cref{sec:structs})
    \item I/O Streams (\cref{sec:streams})
    \item File System abstractions (\cref{sec:fs})
    \item Net Sockets (\cref{sec:sockets})
    \item System and OS abstractions (\cref{sec:system})
\end{itemize}

An object in \eolang{} is either a composition of other objects
  or an atom, which is a foreign function interface
  to a lower-level runtime such as, for example, JVM or Assembly.
In this paper, we denote atoms by a star, for example: \adeff{times}.

Throughout the paper, we use dark blue color
  to highlight the names of objects,
  as opposed to other pieces of fixed-width text.

In this paper%
  \footnote{\raggedright%
    \LaTeX{} sources of this paper are maintained in the
    \href{https://github.com/objectionary/on-the-origin-of-objects}{objectionary/on-the-origin-of-objects} GitHub repository,
    the rendered version is
    \href{https://github.com/objectionary/on-the-origin-of-objects/releases/tag/0.6.0}{\ff{0.6.0}}.}
  we don't provide exact specification of each object,
  in order to avoid conflicts with their exact definitions
  in the Objectionary\footnote{\url{https://github.com/objectionary/home}}.

\section{Principles}

\subsection{Preciseness Over Expressiveness}

We treat objects as the words of an OOP prose:
  the objects that reside at the top level of the root package
  and its sub-packages are the nouns of its vocabulary.
The more concise this vocabulary is and the less ambiguous,
  the cleaner the prose written with it.
Synonyms are especially harmful here,
  since two names for the same meaning make the prose more artistic
  but less strict, forcing the reader to guess which one carries the intent.
We therefore select global objects carefully,
  making sure that none of them duplicate each other,
  and that each delivers a clear and distinct meaning.

\subsection{Transparency Over Speed}

An atom is a function implemented by the runtime beneath \eolang{},
  by the virtual machine or the CPU itself,
  and therefore runs genuinely faster than its \eolang{} counterpart.
An atom is also opaque, though:
  neither the \eolang{} compiler nor any static analyzer
  can look inside it, reason about it, or optimize it.
A program saturated with atoms looks like a solid brick to these tools---
  it either works as it happens to work,
  or nothing can be done about it.
We instead want as many objects as possible to stay transparent to the compiler,
  which we expect to reason about them more powerfully
  than the programmers of atoms ever could.
Eventually, we expect an OO program written in \eolang{},
  resting on only a handful of atoms,
  to be optimized down to highly efficient assembly code.

\subsection{Reusability Over Complexity}

We treat our objects as the building blocks of larger OOP systems,
  and we want every block to be used and to be known to the \eolang{} compiler.
A programmer who reaches for them should never need to reinvent the wheel:
  we intend to supply everything,
  from array sorters and regular expression processors
  to file readers and writers.
The more reusable each object is, the weaker the temptation
  to implement something ``better'' elsewhere,
  and the fewer the private duplicates scattered across programs.
We accept that a few of our objects may grow complex,
  or carry functionality that feels redundant, and that is fine.
Later, once the \eolang{} compiler becomes powerful enough,
  it will know how to reduce that complexity and optimize it away at compile time.

\subsection{Decoration Over Composition}

Decoration is our primary method of extending functionality.
When an object lacks a feature,
  we expect the programmer to build a new object that looks like the original one,
  keeping the same interface but behaving differently.
Wrapping objects into composites with new interfaces and more powerful features,
  as the Adapter and Facade patterns suggest~\citep{gamma1994design},
  is another way to extend what already exists, and we discourage it:
  a new interface hides the original object from its users and from the compiler,
  while a decorator stays interchangeable with the object it decorates.

\subsection{Currying Over Polymorphism}

Suppose an object \ff{books} must tell whether it contains a given title.
The traditional object-oriented answer is polymorphism:
  \ff{books} exposes a \ff{contains} attribute
  and the caller writes \ff{books.contains "Foo"}.
Different kinds of \ff{books} implement it differently---
  an in-memory list scans its elements,
  while a database-backed one runs a \ff{SELECT} query.
The algorithm stays under the control of \ff{books},
  which is the strength of polymorphism,
  but every such feature enlarges the interface of \ff{books},
  and that interface keeps bloating as the code base grows.

The opposite answer moves \ff{contains} out of \ff{books}
  and makes it a free object that takes \ff{books} as an argument,
  so the caller writes \ff{contains books "Foo"}.
Now \ff{books} stays small,
  but polymorphism is lost:
  \ff{books} no longer decides how the search is performed.

We suggest a third answer that keeps both benefits.
The searching object is handed to the caller already bound to its collection,
  as a partially applied object,
  and the caller only completes it, as \ff{contains "Foo"}.
The interface of \ff{books} stays minimal,
  yet the concrete algorithm is still chosen by whoever assembled \ff{contains}.
This is currying, or partial application~\citep{barendregt2012},
  familiar from functional languages such as Haskell,
  brought into an object-oriented setting.

\section{Bytes}\label{sec:bytes}

The center of the taxonomy is the \deff{bytes} object,
  which is an abstraction of a sequence of bytes.
The sequence can either be empty or contain
  a theoretically unlimited number of bytes.
A byte is a unit of information that consists
  of eight adjacent binary digits (bits), each of which consists of a \ff{0}
  or \ff{1}.
In \eolang{} syntax, a sequence of bytes is denoted as either \eohex{---}
  (double dash) for an empty one or as a concatenation of \eohex{xx-},
  where \ff{xx} is a hexadecimal representation of a byte.
For example, the \ff{2A-} is a one-byte sequence,
  where the only byte equals to the decimal number 42.

The \deff{bytes} objects has a few attributes for bitwise operations:
  \adeff{and},
  \adeff{or},
  \adeff{xor},
  \adeff{not},
  \adeff{right}
  and
  \deff{left}.

The object also has the \adeff{eq} attribute for byte-by-byte comparing
  itself with another sequence of bytes.

The \adeff{size} attribute is the total number of bytes in the sequence.

The \adeff{slice} attribute represents a subsequence inside the current one.

The \adeff{concat} attribute is a new sequence,
  which is a concatenation of two byte sequences:
  the current and the provided one.

\section{Boolean}\label{sec:boolean}

The \deff{bool} object abstracts a boolean value, parameterized by
  its \deff{if} attribute---a two-argument object that returns either
  its first argument (for ``true'') or its second one (for ``false'').
The two objects \deff{true} and \deff{false} are simply two applications
  of \deff{bool}, with opposite implementations of \deff{if}.

Every boolean also has \deff{not}, \deff{and}, and \deff{or} attributes.

\section{Number}\label{sec:number}

The object \deff{number} is an abstraction of a floating point number
  of eight bytes according to \citet{ieee754}.
The bytes are encapsulated as the \deff{as-bytes} attribute.

The \deff{eq} attribute is \deff{true} if the \deff{as-bytes}
  is equal to another object, through the use of its \deff{eq}.

The \adeff{plus} attribute is a sum of this number and another \deff{number}.
The \deff{minus} attribute is the subtraction
  of another \deff{number} from this number.
The \adeff{times} attribute and \adeff{div} are multiplication
  and division of this number and another \deff{number}.
The \deff{neg} attribute is the same number with a different sign.
The \deff{floor} attribute is a rounded down \deff{number}.

The \adeff{gt} attribute is \deff{true} if it is ``greater than''
  another \deff{number} and \deff{false} otherwise.
Attributes \deff{lt}, \deff{lte}, and \deff{gte} are ``less than,''
  ``less than or equal,'' and ``greater than or equal''
  comparisons respectively.

The \deff{is-nan}, \deff{is-finite}, and \deff{is-integer} attributes
  are predicates about the number, while \deff{as-i64}, \deff{as-i32},
  and \deff{as-i16} convert it to fixed-width integers.

Special values are provided as separate objects: \deff{nan} for
  ``not a number,'' together with \deff{positive-infinity} and
  \deff{negative-infinity}, also available as the short constants
  \deff{pinf} and \deff{ninf}.

\section{String}\label{sec:string}

The \deff{string} object is an abstraction of a piece of
  Unicode text in UTF-8 encoding, according to \citet{unicode}.
UTF-8 uses one byte for the first 128 code points,
  and up to 4 bytes for other characters.
The text is encapsulated as the \deff{as-bytes} attribute.

The \adeff{slice} attribute takes a piece of a string as another \deff{string}.
The \deff{length} attribute is a total count of characters in the string.

Strings in other encodings, such as UTF-16 or CP1251,
  may be implemented through direct manipulations with \deff{bytes}.

\section{Tuple}\label{sec:tuple}

The \deff{tuple} object is an abstraction of an immutable sequence of objects.
For example, this code represents a tuple \ff{x} of three numbers:

\begin{ffcode}
* 65535 0xF21 3.14 > x
\end{ffcode}

The \deff{with} attribute is a new \deff{tuple} with all elements
  from the current tuple and a new element added to the end of it.
The \deff{at} attribute is the element of the tuple at the specified
  index---positive indices are zero-based starting from the left,
  and negative indices start at -1 from the right.
The \deff{length} attribute is the total number of elements in the tuple.

\section{Error Handling}\label{sec:errors}

The \adeff{error} object terminates the program with an error.
A copy is made with an encapsulated object as the error message;
  the first attempt to dataize it leads to a runtime error and
  program termination.

The \adeff{try} object is the only mechanism able to catch such an error.
It expects three arguments: \deff{main}, \deff{catch}, and \deff{finally}.
When dataized, it first attempts to dataize \deff{main}; if an error
  occurs, the error is caught and \deff{catch} is dataized instead,
  receiving the error as its only argument; the \deff{finally} object is
  dataized afterwards in either case.
This code prints a message instead of crashing on the division by zero:

\begin{ffcode}
try
  10.div 0
  io.stdout "cannot divide" > [e]
  true
\end{ffcode}

\section{Flow Control}\label{sec:flow}

The \deff{seq} object is an abstraction of sequence of objects
  to be dataized sequentially.
For example, this code prints one message to the console
  and then terminates the program due to
  the inability to dataize the division by zero in the middle:

\begin{ffcode}
seq *
  io.stdout "Hello, world!"
  42.div 0
  io.stdout "Bye!"
\end{ffcode}

The objects \deff{true} and \deff{false} both have
  an attribute \deff{if} which is a branching mechanism,
  expecting two arguments: a ``positive'' object, and a ``negative'' object.
When being dataized, the object \ff{true.if} always returns
  the ``positive'' object, the object \ff{false.if} always
  returns the ``negative'' object.
This code gives a title to a random number:

\begin{ffcode}
ms.random.pseudo > r
io.stdout
  tt.sprintf *1
    "Coin toss: 
    if.
      r.gte 0.5
      "heads"
      "tails"
\end{ffcode}

The \deff{while} object is an iterating mechanism, expecting two arguments.
Both are abstract objects with one free attribute for current iteration index.
The first argument is a condition, the second one is the body.
When being dataized, the object \deff{while} repeatedly dataizes
  the body while the \deff{if} of the condition goes ``positive.''
In the end, it returns the body or \deff{false} if no iterations happened.
This code makes a few attempts to find a random number
  that is smaller than 0.1:

\begin{ffcode}
ms.random.pseudo > r
while
  r.gte 0.1 > [i] >>
  io.stdout > [i]
    tt.sprintf *1
      "Attempt #
      i
\end{ffcode}

The \deff{go} object enables forward and backward ``jumps''
  either immediately finishing dataization or restarting it.
For example, this code won't print a message to the console
  if \ff{x} is equal to zero:

\begin{ffcode}
go.to
  seq * > [g]
    if > y
      x.eq 0
      g.forward 0
      42.div x
    io.stdout
      tt.sprintf *1
        "42/x = 
        y
\end{ffcode}

In this example, the number is read from the console again,
  if it is equal to zero:

\begin{ffcode}
go.to
  seq * > [g]
    at. > x
      tt.sscanf
        "
        io.stdin.next-line
      0
    if
      x.eq 0
      g.backward
      io.stdout
        tt.sprintf
          "Number 
          * x
\end{ffcode}

The \deff{switch} object is an abstraction of multi-case
  branching that encapsulates a \deff{tuple} of \((k, v)\) pairs;
  the dataization result is equal to \(v\) of the first pair where
  \(k\) is \deff{true}, or \deff{true} if no pair matches:

\begin{ffcode}
io.stdout
  switch *
    *
      password.eq "swordfish"
      "password is correct!"
    *
      password.eq ""
      "empty password is not allowed"
    *
      false
      "password is wrong"
\end{ffcode}

\section{Digits}\label{sec:digits}

The \deff{i16}, \deff{i32}, and \deff{i64} objects are decorators
  of \deff{bytes} with a predefined size.
They implement the same numeric operations as \deff{number},
  and can be converted to \deff{number} through the \adeff{as-number} attribute.
For example, the number \(42\) as a 32-bit integer always occupies
  exactly four bytes, regardless of its value:

\begin{ffcode}
42.as-i32.as-bytes
\end{ffcode}

The result is \eohex{00-00-00-2A}.

\section{Memory}\label{sec:memory}

The \deff{malloc} object is an abstraction of a block of
  random-access memory.
A copy is created through one of its attributes: \deff{of} allocates
  a block of a given size, \deff{empty} allocates an empty block, and
  \deff{for} allocates a block sized to hold a given object.
The allocated block behaves like \deff{bytes}, but additionally has
  \adeff{write}, \adeff{read}, \deff{copy}, and \deff{put} attributes.
In this example, one kilobyte of memory with random data
  is allocated when a copy of \deff{malloc} is created,
  six bytes are written starting at the 200th byte,
  and then a string is read back and \ff{"Hello!"} is printed:

\begin{ffcode}
malloc.of
  1024
  seq * > [m]
    m.write 200 "Hello!"
    io.stdout
      string
        m.read 200 6
\end{ffcode}

The \deff{malloc} object also supports resizing via
  the \adeff{resized} attribute and copying the data within
  the allocated memory block via the \deff{copy} attribute.
In the next example, \deff{malloc} is allocated and filled with 5 bytes,
  then resized to 10 bytes and its content is duplicated.
The result block contains \ff{"hellohello"}.

\begin{ffcode}
malloc.of
  5
  seq * > [m]
    m.write 0 "hello"
    m.resized 10
    m.copy 0 5 5
\end{ffcode}

The result of dataization of \deff{malloc} is the result
  of dataization of its second argument, which is the scope
  where the memory block is allocated and cleared automatically
  when dataization has happened.

\section{Math}\label{sec:math}

All objects in this Section belong to the \ff{ms} package.
The \deff{random} object is a pseudo-random number generator
  parameterized by a \deff{seed}; when dataized it behaves as a
  \deff{number} between zero and one.
The \deff{next} attribute is the next generator in the sequence,
  the \deff{clocked} attribute seeds the generator from a given clock,
  and the \deff{pseudo} attribute seeds it from the system clock.

The \deff{angle} object is a decorator of \deff{number}.
It assumes that the angle is in radians and has the following attributes
  for trigonometric functions:
  \adeff{sin}, \adeff{cos}, \deff{tan}, and \deff{ctan}.
The \deff{degrees} attribute converts the angle from radians to degrees,
  and the \deff{radians} attribute converts it back.

The constants \deff{pi} and \deff{e} approximate \(\pi\) and Euler's number.
The \deff{real} object is a decorator of \deff{number}.
The \deff{numbers} object decorates a \deff{tuple} of numbers and provides
  \deff{max} and \deff{min} attributes.
The \deff{integral} object calculates the definite integral of a function
  between two limits.
For example, this code approximates the integral of \(f(x) = x\)
  from one to ten with fifteen partitions, yielding approximately \(49.5\):

\begin{ffcode}
ms.integral
  x > [x]
  1
  10
  15
\end{ffcode}

A few more decorators of \deff{number}:

\begin{itemize}
    \item \deff{abs}: absolute value of the number
    \item \deff{exp}: Euler's number raised to the given power
    \item \deff{mod}: remainder of the division by another number
    \item \adeff{sqrt}: square root (\(\sqrt{\rho}\))
    \item \adeff{pow}: the number raised to a given power (\(\rho^x\))
    \item \adeff{ln}: natural logarithm (\(\ln \rho\))
    \item \adeff{acos}: arccosine
    \item \adeff{asin}: arcsine
\end{itemize}

\section{Texts}\label{sec:strings}

All objects presented in this Section belong to the \ff{tt} package.

The \adeff{sscanf} object encapsulates two \deff{string} objects:
  a format and a content.
It behaves like a tuple of scanned data objects.
For example, this code parses a hexadecimal number from the console:

\begin{ffcode}
at.
  tt.sscanf
    "
    io.stdin
  0
\end{ffcode}

The \adeff{sprintf} object builds a string according to the format
  provided and compliant with \citet[Chapter 5]{posix}, for example:

\begin{ffcode}
tt.sprintf *1
  "Hi, 
  "Jeff"
  3.14
\end{ffcode}

A few more decorators of \deff{string}:

\begin{itemize}
    \item \deff{trimmed}: removes both leading and trailing spaces
    \item \deff{trimmed-left}: removes leading spaces
    \item \deff{trimmed-right}: removes trailing spaces
    \item \deff{joined}: joins a tuple of strings with this one as a glue
    \item \deff{chained}: concatenates itself with other strings
    \item \deff{repeated}: repeats itself a given number of times
    \item \deff{slice}: takes a piece of itself as another string
    \item \deff{at}: retrieves the symbol at a given index
    \item \deff{contains}: checks whether it contains a substring
    \item \deff{contains-all}: checks whether it contains all given substrings
    \item \deff{contains-any}: checks whether it contains any given substring
    \item \deff{starts-with}: checks whether it starts with a substring
    \item \deff{ends-with}: checks whether it ends with a substring
    \item \deff{index-of}: finds the first occurrence of a substring
    \item \deff{last-index-of}: finds the last occurrence of a substring
    \item \deff{up-cased}: makes it upper case
    \item \deff{low-cased}: makes it lower case
    \item \deff{replaced}: finds and replaces a substring
    \item \deff{split}: breaks it into a tuple of strings
    \item \deff{nsplit}: breaks it into at most a given number of strings
    \item \deff{as-number}: converts itself to \deff{number}
    \item \deff{as-ascii}: reads a single character as its ASCII code
    \item \deff{is-alpha}: checks whether all characters are letters
    \item \deff{is-digit}: checks whether all characters are digits
    \item \deff{is-alphanumeric}: checks whether all characters are letters or digits
    \item \deff{is-ascii}: checks whether all characters are ASCII
\end{itemize}

The \deff{string-buffer} object allows building a string incrementally,
  appending pieces through its \deff{with} attribute.
For example, this code builds the string \ff{"Hello, Jeff"}
  and afterwards behaves as a plain \deff{string}:

\begin{ffcode}
tt.string-buffer "Hello"
  .with ", "
  .with "Jeff"
\end{ffcode}

The \deff{regex} object is an abstraction of a regular expression
  in Perl-compatible format, with full support of Unicode.
The \deff{matches} attribute is \deff{true} if the given text
  matches the pattern.
The \deff{match} attribute is a matcher over the given text: its
  \deff{next} attribute is the first matched block, each block exposes
  \deff{text}, \deff{start}, and \deff{group}, and its own \deff{next}
  attribute is the following block.

\begin{ffcode}
tt.regex "/[a-z]+/" > r
r.match "!hello!world!" > m
eq.
  m.next.text
  "hello"
\end{ffcode}

\section{Structs}\label{sec:structs}

All objects presented in this Section belong to the \ff{ss} package.

The object \deff{list} is a decorator of \deff{tuple} that adds a
  collection API.
The attribute \deff{is-empty} is \deff{true} if the length of the tuple is zero.
The attribute \deff{eq} is \deff{true} if both lists have the same length
  and each element is equal to the corresponding element of another list.
The attribute \deff{with} is a new list with a new element
  appended to the end of it, while \deff{withi} inserts a new element
  at the \(i\)-th position.
The attribute \deff{without} is a new list with all elements equal
  to the given one removed.
The attribute \deff{each} dataizes a given object for every element.
The \deff{front} and \deff{back} attributes are the first and the last
  \(N\) elements, \deff{slice} is a sub-list between two indices, and
  \deff{shifted} is the list without its first \(N\) elements.
The \deff{sorted} attribute is a new list with elements sorted using
  the \deff{lt} attribute of the elements.
The \deff{concat} attribute is a new list that concatenates
  the current list with the one provided as an argument.

The functional operations over a list are standalone objects of the
  \ff{ss} package, each taking the list as its first argument.
The objects \deff{reduced}, \deff{mapped}, \deff{filtered}, and
  \deff{eachi} are respectively similar to \ff{reduce()}, \ff{map()},
  \ff{filter()}, and \ff{forEach()} methods of the \ff{Array} object
  in JavaScript~\citep{EcmaScript}.
The ``twin'' objects \deff{reducedi}, \deff{mappedi}, and \deff{filteredi}
  are semantically the same, but pass an extra \deff{number} index to their
  function.
The \deff{contains}, \deff{index-of}, and \deff{last-index-of} objects
  search a list for an element, while \deff{withouti} is a new list with
  the \(i\)-th element removed.
The \deff{bytes-as-array} object represents a sequence of \deff{bytes}
  as a \deff{tuple} of one-byte sequences.
For example, this code doubles every element of a list,
  keeps the ones greater than four,
  and sums the survivors, resulting in \(14\):

\begin{ffcode}
ss.reduced
  ss.filtered
    ss.mapped
      ss.list (* 1 2 3 4)
      x.times 2 > [x]
    x.gt 4 > [x]
  0
  a.plus x > [a x]
\end{ffcode}

The object \deff{map} is a hash map built from a \deff{tuple} of \deff{map.entry}
  objects which are pairs \((k, v)\).
The \deff{map} ensures that all \(k\) are always unique.
It's expected that each \(k\) is dataizable so it's possible to calculate \deff{hash-code-of} it.
The \deff{with} attribute is a new map with a pair added or replaced,
  and \deff{without} is a new map with a key removed.
The \deff{found} attribute tries to find an object in the map by the given key.
The returned object has two attributes:
  \deff{exists} which is \deff{true} if the object was found;
  and \deff{get} which is the found object or \adeff{error}.
The \deff{keys} attribute is the \deff{list} of map keys.
The \deff{values} attribute is the \deff{list} of map values.
The \deff{size} attribute is the \deff{size} of the map.
The \deff{has} attribute is \deff{true} if the map contains the given \(key\).
For example, this code builds a map of two entries
  and reads the value \(2\) back through its key:

\begin{ffcode}
ss.map > m
  *
    ss.map.entry "one" 1
    ss.map.entry "two" 2
(m.found "two").get
\end{ffcode}

The object \deff{set} is an unordered collection of unique objects,
  built on top of \deff{map}.
It's expected that each element is dataizable so it's possible to calculate \deff{hash-code-of} it.
Its \deff{with} and \deff{without} attributes add and remove an element,
  \deff{has} checks membership, and \deff{size} is the number of elements.

The object \deff{hash-code-of} is the pseudo-unique floating-point numeric
  representation of an object.
It's expected that the given object is dataizable.

The object \deff{range} is a \deff{list} that contains a range of elements.
It accepts two attributes: \deff{start} and \deff{end}.
The attribute \deff{start} must be an abstract object that must have an attribute
  \deff{next} to get the next element and an attribute \deff{lt} to compare with
  the \deff{end} object.
Every next element also must have attributes \deff{next} and \deff{lt}.
The first object in the chain must have a default value.

The object \deff{range-of-numbers} is a \deff{range} of integer numbers
  from \deff{start} to \deff{end} (soft border) with step = \(1\).
It accepts two void attributes \deff{start} and \deff{end} which must be integer \deff{number}s.
If they're not, an \adeff{error} is returned.
For example, \ff{ss.range-of-numbers 1 10} is the list
  \ff{* 1 2 3 4 5 6 7 8 9}, since the upper bound is excluded.

\section{I/O Streams}\label{sec:streams}

All objects presented in this Section belong to the \ff{io} package.

It is expected that an input stream has the following interface:

\begin{ffcode}
[] > input
  [size] > read /bytes
\end{ffcode}

It is expected that an output stream has the following interface:

\begin{ffcode}
[] > output
  [bytes] > write /true
\end{ffcode}

The \deff{console} object is a basic I/O object that allows interaction
  with the operating system console.
It behaves as ``input'' and ``output.''
The \deff{stdout} object is an ``output'' that prints a \deff{string} to the standard output stream.
The \deff{stdin} object is an ``input'' and an abstraction of a \deff{string}
  currently available in the standard input stream.
The attribute \deff{next-line} reads a line until the EOL character (\ff{\char`\\n}).
The attribute \deff{all-lines} reads all lines as long as possible.
If there is no string in the stream, the object blocks dataization and waits.
This code endlessly reads strings from the console and immediately prints them back:

\begin{ffcode}
while
  true > [i]
  io.stdout > [i]
    io.stdin.next-line
\end{ffcode}

The object \deff{bytes-as-input} is an ``input'' created from \deff{bytes}.
The object \deff{malloc-as-output} is an ``output'' directed towards a copy of \deff{malloc}.
The object \deff{tee-input} is an ``input'' and a channel between an ``input''
  and an ``output,'' which moves all available bytes from the former to the
  latter when being dataized.
For example, this code writes text to a temporary file:

\begin{ffcode}
(fs.file "/tmp/foo.txt").open
  [f]
    io.tee-input
      io.bytes-as-input
        "Hello, world!".as-bytes
      f
\end{ffcode}

The object \deff{input-length} is an object which reads all the bytes
  from the provided ``input'' and returns its length.
The object \deff{input-as-bytes} reads all the bytes from the provided ``input''
  and returns them as \deff{bytes}.
The object \deff{copied} copies all the bytes from an ``input'' to an ``output''
  and returns the number of bytes copied.
The objects \deff{dead-input} and \deff{dead-output} are stubs which read from and write to nothing.

\section{File System}\label{sec:fs}

Here we define manipulations with files and directories.
All objects presented in this Section belong to \ff{fs} package.

The object \deff{path} is an abstraction of file path and is a decorator of \deff{string}.
The attribute \deff{separator} is a system-dependent default name-separator character.
The attribute \deff{joined} is a utility object that joins the given \deff{tuple}
  of paths with the current OS separator and normalizes the result path.
The attribute \deff{as-file} is a \deff{file} created from the path itself.
The attribute \deff{as-dir} is a \deff{dir} created from the path itself.
The attribute \deff{is-absolute} is \deff{true} if the path is absolute and \deff{false} otherwise.
The attribute \deff{normalized} is a new \deff{path} with normalized \(uri\).
Normalization includes converting multiple slashes into a single slash
  and resolving ``.'' (current directory) and ``..'' (parent directory) segments.
The attribute \deff{resolved} is a new \deff{path} with a suffix appended to the current one.
The attribute \deff{basename} is the name of the file in the path.
The attribute \deff{extname} is the extension in the path.
The attribute \deff{dirname} is the name of the directory in the path.

The object \deff{file} is an abstraction of a file and is a decorator of \deff{path},
  which is the path of the file.
The attribute \deff{as-path} is a \deff{path} of the file.
The attribute \adeff{is-directory} is \deff{true} if the file is a directory.
The attribute \adeff{exists} is \deff{true} if the file exists.
The attribute \deff{touched} makes sure the file exists.
The attribute \deff{deleted} removes the file.
The attribute \adeff{size} is the size of the file in bytes.
The attribute \deff{moved} renames or moves the file to a new place.
The attribute \deff{open} opens the file.
The \deff{open} object accepts two attributes: file open mode which is a \deff{string}
  and file scope.
The file scope attribute must be an abstract object with one void attribute
  which is \deff{file-stream}.
It has the attributes \adeff{read} and \adeff{write} which allow working with the file
  as ``input'' and ``output''.
The result of dataization of \ff{file.open} is the result of dataization
  of its second argument ``scope''.
The file descriptor is closed automatically when the scope is dataized.

The object \deff{dir} is an abstraction of a directory.
The attribute \deff{made} creates the directory with all its parent directories.
The attribute \deff{deleted} deletes the directory recursively with all its subdirectories.
The attribute \adeff{walk} finds files in the directory using glob pattern matching.
The attribute \deff{tmpfile} is a temporary file in the directory.

The object \deff{tmpdir} is a decorator of \deff{dir} that abstracts the system
  directory for temporary files.

In the next example, a temporary file is created and filled up with \ff{"Hello, world"}.
Then the data is read from the file and written to \deff{malloc}.

\begin{ffcode}
malloc.of
  12
  seq > [m]
    *
      tmpdir
      .tmpfile
      .open
        "w"
        f.write "Hello, world" > [f]
      .open
        "r"
        m.put > [f] >>
          f.read 12
      m
\end{ffcode}

\section{Net}\label{sec:sockets}

Objects used for both server-side and client-side TCP sockets
  work in accordance with \citet{posix}.
All objects presented in this Section belong to the \ff{nk} package.

The \deff{socket} object is an abstraction of a TCP socket.
The \deff{connect} attribute establishes a connection.
It accepts an abstract object with one void attribute which is the socket descriptor.
The socket descriptor has the following attributes:

\begin{itemize}
    \item \deff{send}: send a message to the socket
    \item \deff{recv}: receive \(x\) amount of bytes from the socket
    \item \deff{as-input}: make ``input'' from socket
    \item \deff{as-output}: make ``output'' from socket
\end{itemize}

When you dataize \ff{socket.connect}, it creates a new socket,
  connects to the given IP on the given port, dataizes the provided scope,
  closes the socket and returns bytes retrieved from scope dataization.
An \adeff{error} is returned if any of the described steps failed.

This code opens a connection to TCP port 80 of \ff{localhost} and
  then reads the entire stream as a Unicode string:

\begin{ffcode}
(nk.socket "127.0.0.1" 80).connect
  io.tee-input > [s]
    s.as-input
    io.console
\end{ffcode}

The \deff{listen} attribute allows you to listen to incoming connections as a server.
It accepts an abstract object with one void attribute which is the socket descriptor.
The current socket descriptor has the same attributes as the descriptor
  from \ff{socket.connect} and also one extra attribute \deff{accept}.
This attribute is used to wait for an incoming connection.
It also accepts an abstract object with one void attribute
  which is the client socket descriptor, which was connected to your server.

This code emulates a server on TCP port 8080 of \ff{localhost},
  then it accepts an incoming connection and sends \ff{"Hello, world"} to it:

\begin{ffcode}
(nk.socket "127.0.0.1" 8080).listen
  s.accept > [s]
    client.send "Hello, world" > [client]
\end{ffcode}

\section{System}\label{sec:system}

All objects presented in this Section belong to the \ff{sm} package.
They provide access to the underlying operating system.

The \deff{os} object represents the current operating system.
The \adeff{name} attribute is its name, while \deff{is-windows},
  \deff{is-linux}, and \deff{is-macos} are predicates about its family.

The \deff{getenv} object reads an environment variable as a \deff{string};
  an empty string means the variable does not exist.

The \deff{eol} object, also available as \deff{line-separator}, is the
  system-dependent line separator: \ff{\char`\\n} on UNIX
  and \ff{\char`\\r\char`\\n} on Windows.

The \deff{system-clock} object is a wall-clock time source backed by the
  operating system; it decorates an \deff{i64} holding the number of
  milliseconds since the Unix epoch.

The \deff{posix} object makes a Unix system call by name through the POSIX
  interface, while the \deff{win32} object makes a \ff{kernel32.dll}
  function call by name; both accept a \deff{tuple} of arguments.
The call returns an object whose \deff{code} attribute is the numeric
  return code of the syscall and whose \deff{output} attribute is its output.
For example, this code invokes the \ff{getpid} syscall with an empty
  \deff{tuple} of arguments and reads back the identifier
  of the currently running process:

\begin{ffcode}
(sm.posix "getpid" tuple.empty).code
\end{ffcode}

\raggedright
\bibliographystyle{ACM-Reference-Format}
\bibliography{bibliography/main}

\end{document}